# 基于数据驱动的燃煤锅炉 NOx 排放量动态修正预测模型

唐振浩[1]，朱得宇[1]，李扬[2]

（1. 东北电力大学自动化工程学院，吉林省 吉林市 132012； 2. 东北电力大学电气工程学院，吉林省吉林市 132012）

# Data Driven based Dynamic Correction Prediction Model for NOx Emission of Coal Fired Boiler

Tang Zhenhao[1]，Zhu Deyu[1]， Li Yang[2]

(1.School of Automation Engineering，Northeast Electric Power University，Jilin Province，China；2. School of Electrical Engineering, Northeast Electric Power University, Jilin 132012, Jilin Province, China)

**ABSTRACT:** The real-time prediction of NOx emissions is of great significance for pollutant emission control and unit operation of coal-fired power plants. Aiming at dealing with the large time delay and strong nonlinear characteristics of the combustion process, a dynamic correction prediction model considering the time delay is proposed. First, the maximum information coefficient (MIC) is used to calculate the delay time between related parameters and NOx emissions, and the modeling data set is reconstructed; then, an adaptive feature selection algorithm based on Lasso and ReliefF is constructed to filter out the high correlation with NOx emissions. Parameters; Finally, an extreme learning machine (ELM) model combined with error correction was established to achieve the purpose of dynamically predicting the concentration of nitrogen oxides. Experimental results based on actual data show that the same variable has different delay times under load conditions such as rising, falling, and steady; and there are differences in model characteristic variables under different load conditions; dynamic error correction strategies effectively improve modeling accuracy; proposed The prediction error of the algorithm under different working conditions is less than 2%, which can accurately predict the NOx concentration at the combustion outlet, and provide guidance for NOx emission monitoring and combustion process optimization.

基金项目：国家自然科学基金(61503072)，吉林省科技发展计划项目(20190201095JC, 20200401085GX)。

The National Natural Science Foundation of China (Grant No.: 61503072) and Jilin Science and Technology Project (Grant No.: 20190201095JC, 20200401085GX)

**摘要**：NOx 排放量实时预测对于燃煤电厂污染物排放控制和机组运行具有重要意义。为了克服燃烧过程大时延及强非线性特性，提出了一种考虑时间延迟的动态修正预测模型。首先，利用最大信息系数(MIC)计算相关参数与 NOx 排放量的延迟时间，重构建模数据集；然后，构建基于 Lasso 和 ReliefF 的自适应特征选择算法，筛选与 NOx 排放量相关程度高的参数；最后，建立了结合误差校正的极限学习机（ELM）模型，达到动态预测氮氧化物浓度的目的。基于实际数据的实验结果表明：相同变量在升、降、平稳等负荷工况下的延迟时间不同；且不同负荷工况下模型特征变量存在差异；动态误差校正策略有效提升建模精度；所提出算法在不同工况下的预测误差均小于 2%，能够准确预测燃烧出口的 NOx 浓度，为 NOx 排放监测和燃烧过程优化提供指导。

## 0 引言

根据国家能源局统计，2020 年全国火力发电量为 52798.7 亿 KW·h，占全国总发电量的 71.2%。虽然我国新能源发展迅速，但火力发电仍是我国现阶段电力来源的主要形式[1]。然而，燃煤电厂产生的 NOx 会污染大气环境，导致酸雨，并严重影响人体健康[2]。因此，国家环保部门对火力发电厂的 NOx 排放要求日益严格[3-4]，目前执行的火电厂 NOx 排放国家标准在 50mg/m$^3$ 以下[5]。NOx 排放量检测主要依靠连续排放监测系统(continuous emission monitoring system，CEMS)[6]，但是，部分电厂使用煤质差，导致燃

烧排放烟气湿度大、灰尘多、腐蚀性强，从而CEMS工作环境恶劣，需定时清灰且易发生故障，同时，CEMS不能对未来NOx排放量进行预测。因此探究NOx排放量软测量方法能够在不增加硬件成本的基础上提高检测稳定性，并对NOx排放量未来变化趋势进行预测。

燃煤锅炉燃烧过程机理复杂[7]，NOx排放量受到机组负荷、配风方式等多种因素共同影响，难以通过传统的机理模型进行计算[8]。随着大量生产数据的积累和数据驱动技术的发展，神经网络、深度学习等数据驱动算法在非线性工业过程建模问题中取得了良好的应用效果[9-11]。在燃煤锅炉NOx排放量预测建模方面，TAN等[12]采用长短期记忆(long short-term memory，LSTM)模型来预测燃煤锅炉的NOx排放量。周昊等[13]应用核心向量机(core vector machine，CVM)通过77个运行参数建立了超超临界锅炉的NOx排放特性模型。Tang等[14]基于深度置信网络(deep belief network，DBN)建立了NOx排放量的预测模型。上述研究证明了数据驱动方法在NOx排放量预测中的可行性。ELM是2004年提出的新型神经网络算法[15]，具有结构简单，训练速度快等特点，因此，本研究采用ELM作为构建特征信息与NOx排放量非线性关系模型的基础算法。

在机组实际运行过程中，由于NOx检测滞后、锅炉燃煤传送滞后、燃烧反应消耗时间等因素，锅炉燃烧过程具有延迟特性。因此，计算各输入特征变量相对于NOx排放量的延迟时间是实现NOx浓度动态预测、提高预测准确度的关键。闫来清等[16]通过k-近邻互信息法计算各输入变量的时间延迟，实现对NOx排放量的动态预测建模。赵征等[17]利用互信息计算各输入变量的延迟时间，将含有延迟信息的输入变量与输出变量进行数据重构，实现对NOx排放量的动态预测。最大信息系数 (maximal information coefficient，MIC)适用于计算大数据下广泛的非线性关系，鲁棒性强且计算复杂度低。为深入挖掘输入变量延迟与NOx排放量之间非线性关系，本研究采用MIC方法表征燃烧过程相关变量不同延迟时间数据与NOx排放量数据之间的非线性相关性。并使用MIC计算不同工况下（升负荷、降负荷、负荷稳定）各相关参数相对于NOx排放量的延迟时间，分别构建预测模型，提高模型精度。

机组运行的历史数据通常存在信息耦合以及变量规模大的问题。高维输入变量增加模型复杂度以及计算时间，导致模型的计算效率降低。因此，输入变量的选择也是NOx排放量预测模型研究的重点。Wang等[18]通过互信息(mutual information，MI)特征选择的方法对输入参数进行相关性计算，得到模型的最佳输入参数。吕游等[19]利用(partial least squares，PLS)对输入变量进行特征提取，降低输入维数，建立NOx排放预测模型。但上述方法仅通过一种特征选择的方法来获取模型的输入，可能导致特征变量的少选或漏选，影响模型的预测精度。因此，本研究综合基于Lasso和ReliefF两种特征选择算法设计自适应特征选择方案，并结合机理分析的结果，为预测模型确定合理的输入特征集合。

为了进一步提升预测精度，在建模方案中引入误差修正策略。Yang等[20]针对NOx排放量建立预测模型时，对预测误差进行建模预测，进一步缩小NOx测量值和预测值之间的偏差。Ouyang等[21]根据初始风功率预测模型的误差建立误差修正模型，提升了风功率预测精度。因此，尝试将误差修正策略引入NOx排放量建模预测中，进一步缩小初始预测模型的预测值和测量值之间的误差。

综上所述，本研究使用最大信息系数MIC计算模型的各输入变量相对输出变量的时间延迟，进行建模数据的重构，以消除燃烧过程中的各参数延迟对预测结果的影响；然后基于Lasso和ReliefF设计自适应的组合特征选择算法，为NOx排放量筛选相关性更大的变量。最后使用极限学习机ELM建立初始NOx排放量预测模型，对预测结果进行误差修正，得到燃煤锅炉NOx动态修正预测模型。

## 1 锅炉燃烧过程简介

本研究对象为1000MW超超临界燃煤机组的$\pi$型锅炉。锅炉的炉膛尺寸为32.084m×15.67m，通过八角反向切圆的燃烧器摆放方式使炉膛内部进行充分燃烧，其中每个燃烧器共设有6层一次风喷口、6层二次风喷口及2层燃尽风喷口。锅炉剩余主要结构包括过热器、再热器、省煤器、空预器等。锅炉燃烧系统如图1所示，煤经给煤机运送至磨煤机，空预器将冷空气加热为热空气，热空气分为一次风和二次风，

一次风将煤粉吹入炉膛，二次风补充煤粉继续燃烧所需要的空气。锅炉炉膛内部燃烧产生的烟气经炉膛至过热器，后经省煤器出口流出。并在图1中标注了建模使用的关键锅炉变量。

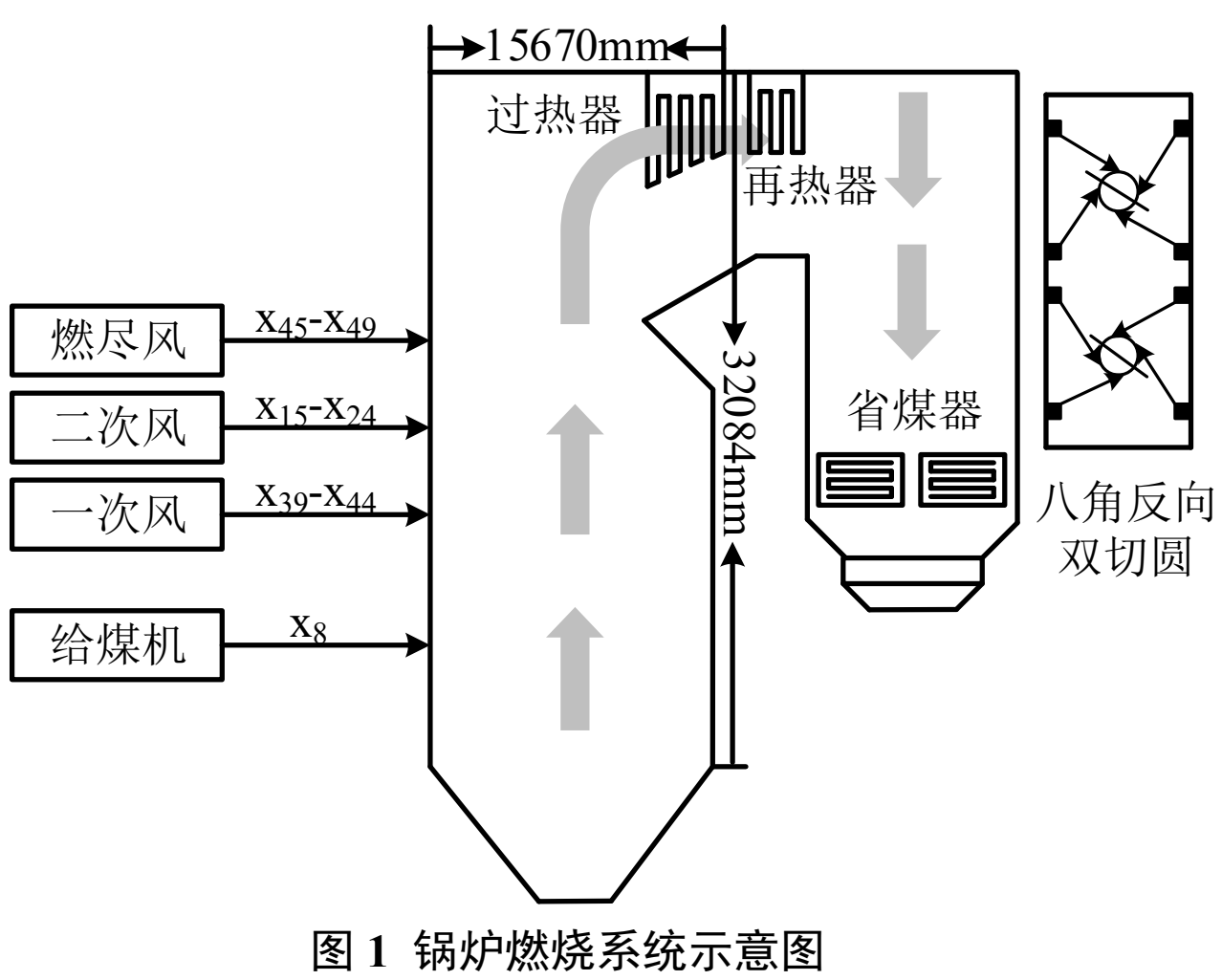

**图 1 锅炉燃烧系统示意图**

**Fig. 1 The schematic diagram of the boiler combustion system**

数据来自电厂的分散控制系统（distributed control system，DCS）中，采样间隔为5s。根据NOx产生机理分析，选择主蒸汽压力、主蒸汽温度、总风量、氧量、总煤量、一次风风量、二次风风量、给煤量、机组负荷、燃尽风、锅炉尾部烟道烟气氧量等55个变量作为与产生NOx相关的变量。由于电厂无法实时在线获取燃烧中的煤质参数，因此在建立NOx预测模型时无法将煤质特性作为输入参数，并且在建模数据的时间范围内，采用的煤质参数保持恒定。

## 2 NOx 排放预测建模

### 2.1 数据预处理

实验数据为电站DCS系统导出的历史运行数据。为了使模型适应现场运行工况的变化，在建立预测模型时建模数据应尽可能包含更多机组负荷分布情况，分别采用负荷上升、下降、平稳等三种机组负荷分布数据集进行建模。相关变量及变化范围如表1所示。

由于所采集数据中含有异常值，影响实验的准确性，采用拉依达准则进行异常值的筛选，并使用异常值前10个采样点数据平均值对筛选出来的异常值进行替代。电站锅炉中各个运行参数的取值范围差距较大，取值范围小的参数对NOx排放量的影响容易被忽略，影响模型精度。因此对所有变量进行归一化处理，将数据统一到[0，1]区间内，使变量处于同一量级，计算方法如式(1)所示：

$$x' = \frac{x - x_{\min}}{x_{\max} - x_{\min}} \tag{1}$$

其中$x'$表示变量归一化后的数据，$x$表示变量的原始测量值，$x_{\min}$表示变量的最小值，$x_{\max}$表示变量的最大值。

**表 1 在运行条件下所研究锅炉的部分过程变量**

**Tab. 1 Part of process variables of the investigated boiler in the operation condition**

| 变量 | 标签 | 单位 | 变量范围 |
| --- | --- | --- | --- |
| 主蒸汽压力 | *x*1 | MPa | [16.771-31.379] |
| 主蒸汽温度 | *x*2 | ℃ | [571.708-605.354] |
| 水煤比 | *x*3 | — | [6.617-8.941] |
| 过热度 | *x*4 | ℃ | [34.053-77.287] |
| 总风量 | *x*5 | t/h | [1798.56-3187.22] |
| 氧量 | *x*6 | % | [2.596-5.647] |
| 排烟温度 | *x*7 | ℃ | [130.071-140.147] |
| 总煤量 | *x*8 | t/h | [158.303-322.039] |
| 磨煤机出口风温(A-F) | *x*9-*x*14 | ℃ | [21.923-86.578] |
| 二次风风量(A 层-F 层) | *x*15-*x*24 | t/h | [34.303-209.429] |
| 机组负荷 | *x*25 | MW | [494.647-999.145] |
| 炉膛压力 | *x*26 | kPa | [-0.341-0.142] |
| 给煤机瞬时煤量(A-F) | *x*27-*x*32 | t/h | [38.864-76.959] |
| 磨煤机进口一次风温(A-F) | *x*33-*x*38 | ℃ | [29.446-218.082] |
| 磨煤机进口一次风量(A-F) | *x*39-*x*44 | t/h | [98.142-164.633] |
| 下层燃尽风流量(B-D) | *x*45-*x*47 | t/h | [39.692-132.824] |
| 上层燃尽风流量 | *x*48-*x*49 | t/h | [34.019-141.758] |
| 一次再热器温度 | *x*50 | ℃ | [596.977-605.507] |
| 二次再热器温度 | *x*51 | ℃ | [567.125-606.765] |
| 锅炉尾部烟气氧量(1-4) | *x*52-*x*55 | % | [2.288-6.096] |

### 2.2 确定延迟时间

由于锅炉在燃烧过程中各输入变量参与燃烧

反应的时间不同、传感器测量的延迟等[22]，导致燃烧产生的 NOx 排放量相对其余运行参数存在延迟时间。DCS 系统记录的同一时间标签下运行数据中的各相关参数之间未形成准确对应关系。若不考虑各输入变量与输出 NOx 的延迟时间，则无法准确建立 NOx 排放量与输入变量之间准确的非线性动态映射关系。因此，需要确定输入变量与输出变量之间的延迟时间，根据计算结果及机理分析指导建模数据重构。

**表 2 不同运行工况下各辅助参数与 NOx 浓度之间的延迟时间**

**Tab. 2 Delay time between auxiliary parameters and NOx concentration under different operating conditions**

| 参数 | 延迟时间（s） | | | 参数 | 延迟时间（s） | | |
|---|---|---|---|---|---|---|---|
| | 升 | 降 | 稳 | | 升 | 降 | 稳 |
| *x*1 | 10 | 0 | 0 | *x*29 | 120 | 115 | 125 |
| *x*2 | 145 | 110 | 75 | *x*30 | 130 | 110 | 80 |
| *x*3 | 95 | 110 | 80 | *x*31 | / | / | 50 |
| *x*4 | 20 | 10 | 50 | *x*32 | / | 120 | 115 |
| *x*5 | 85 | 85 | 90 | *x*33 | 55 | 40 | 100 |
| *x*6 | 0 | 0 | 0 | *x*34 | 45 | 35 | 20 |
| *x*7 | 210 | 200 | 230 | *x*35 | 90 | 15 | 55 |
| *x*8 | 115 | 115 | 110 | *x*36 | 50 | 15 | 20 |
| *x*9 | 215 | 260 | 235 | *x*37 | 45 | 45 | 60 |
| *x*10 | 180 | 175 | 200 | *x*38 | 80 | 85 | 95 |
| *x*11 | 110 | 250 | 170 | *x*39 | / | / | 0 |
| *x*12 | 230 | 280 | 260 | *x*40 | 10 | 40 | 20 |
| *x*13 | 185 | 210 | 285 | *x*41 | 40 | 35 | 40 |
| *x*14 | 190 | 170 | 80 | *x*42 | 30 | 25 | 30 |
| *x*15 | 165 | 150 | 100 | *x*43 | / | / | 30 |
| *x*16 | 200 | 190 | 220 | *x*44 | 0 | 0 | 0 |
| *x*17 | 205 | 160 | 175 | *x*45 | 50 | 40 | 65 |
| *x*18 | 200 | 175 | 245 | *x*46 | 80 | 75 | 60 |
| *x*19 | 295 | 240 | 265 | *x*47 | 85 | 90 | 75 |
| *x*20 | 270 | 290 | 250 | *x*48 | 165 | 110 | 160 |
| *x*21 | 205 | 140 | 100 | *x*49 | 160 | 95 | 150 |
| *x*22 | 290 | 235 | 240 | *x*50 | 160 | 125 | 170 |
| *x*23 | 5 | 5 | 35 | *x*51 | 125 | 110 | 190 |
| *x*24 | 270 | 260 | 295 | *x*52 | 0 | 0 | 0 |
| *x*25 | 70 | 45 | 115 | *x*53 | 0 | 0 | 0 |
| *x*26 | 105 | 60 | 200 | *x*54 | 0 | 0 | 0 |
| *x*27 | / | / | 0 | *x*55 | 0 | 0 | 0 |
| *x*28 | 125 | 120 | 240 | | | | |

为了表征输入变量不同时延重构序列与 NOx 排放量数据的非线性相关性,采用 MIC 方法计算其非线性相关程度[23]，具体计算如公式（2）和（3）所示：

$$MIC(x,y)=\max_{xy<B(n)}\frac{I(x,y)}{\log_2(\min(x,y))} \tag{2}$$

$$I(x,y)=\sum_{x\in X}\sum_{y\in Y}p(x,y)\log(\frac{p(x,y)}{p(x)p(y)}) \tag{3}$$

其中 $I(x,y)$ 为变量 $x$ 和 $y$ 之间的互信息值，$p(x)$、$p(y)$ 为边缘概率分布，$p(x,y)$ 为联合概率分布，$B(n)$ 设置为数据量 $n$ 的 0.6 次方左右。根据现场经验，锅炉燃烧时各变量的延迟时间一般在 0—300s 内，这些变量分别为总风量、氧量、排烟温度、总煤量、磨煤机出口风粉温度、二次风风量（A—F）、磨煤机进口一次风温度（A—F）、锅炉尾部烟道烟气氧量（4 个）等。为了计算每个变量的延迟时间，在 0—300s 内，以 5s 为间隔依次进行该变量的时间序列延迟重构，即相对于 NOx 排放量依次进行 5s 延迟，并加入该变量的当前时刻，共同与输出 NOx 进行 MIC 值求解，选择对应 MIC 值最大的一组重构序列，求得每个变量的最佳的延迟时间。各工况的参数延迟时间如表 2 所示，表 2 中空格部分表示该变量在当前工况未工作状态。从表 2 中可以看出，同一个参数在不同工况下的延迟时间不相同。

### 2.3 特征选取

建立准确的预测模型首先要选取合理的特征变量，减少输入变量的维数，提高预测模型的精度。然而由于燃烧过程的反应复杂、运行工况多变，仅通过机理分析无法完全获取影响 NOx 排放量的特征集合，因此通过特征选择的方法选取与输出 NOx 相关性较大的变量，剔除冗余变量。

Lasso 回归是一种以缩小变量集为思想的压缩估计方法，通过构造一个惩罚函数，可以将对输出变量影响不明显的输入变量的系数压缩至 0，达到变量选择的目的。Lasso 回归是在损失函数后加入 L1 正则化，损失函数如公式(4)所示：

$$J(\theta)=\frac{1}{2n}\sum_{i=1}^{n}(h_\theta(x^i)-y^i)^2 \tag{4}$$

加入 L1 正则化项后如公式(5)所示：

$$J(\theta)=\frac{1}{2n}\sum_{i=1}^{n}(h_\theta(x^i)-y^i)^2+\lambda\sum_{j=1}^{p}\left|\theta_j\right| \tag{5}$$

Lasso 的复杂程度由惩罚项系数 $\lambda$ 来控制，$\lambda$ 越大对变量较多的模型的惩罚力度就越大，从而获得一个变量较少的模型。但 Lasso 回归容易漏选一些对输入影响较大的变量和引入一些无关变量，因此结合 ReliefF 回归算法进一步筛选特征变量。

ReliefF 算法基本内容为：从训练集中随机选择一个样本 $R$，从 $R$ 的同类样本集中取出 $k$ 个近邻样本（near Hits），然后从其他不同类样本集中分别找出 $k$ 个近邻样本，更新每个特征的权重，如公式(6)所示：

$$W(A)=W(A)-\sum_{j=1}^{k}\frac{diff(A,R,H_j)}{mk}+\sum_{C\notin class(R)}\frac{\frac{p(C)}{1-p(class(R))}\sum_{j=1}^{k}diff(A,R,M_j(C))}{mk} \tag{6}$$

式中：$p(C)$ 为该类别的比例，$p(class(R))$ 为随机选取的某样本的类别的比例。$diff(A_1,R_1,R_2)$ 表示样本 $R_1$ 和 $R_2$ 在特征 $A$ 上的差，$M_j(C)$ 表示类 $C\notin class(R)$ 中第 $j$ 个最近邻样本。

基于 Lasso 和 ReliefF 的自适应特征选择算法的主要流程如图 2 所示：

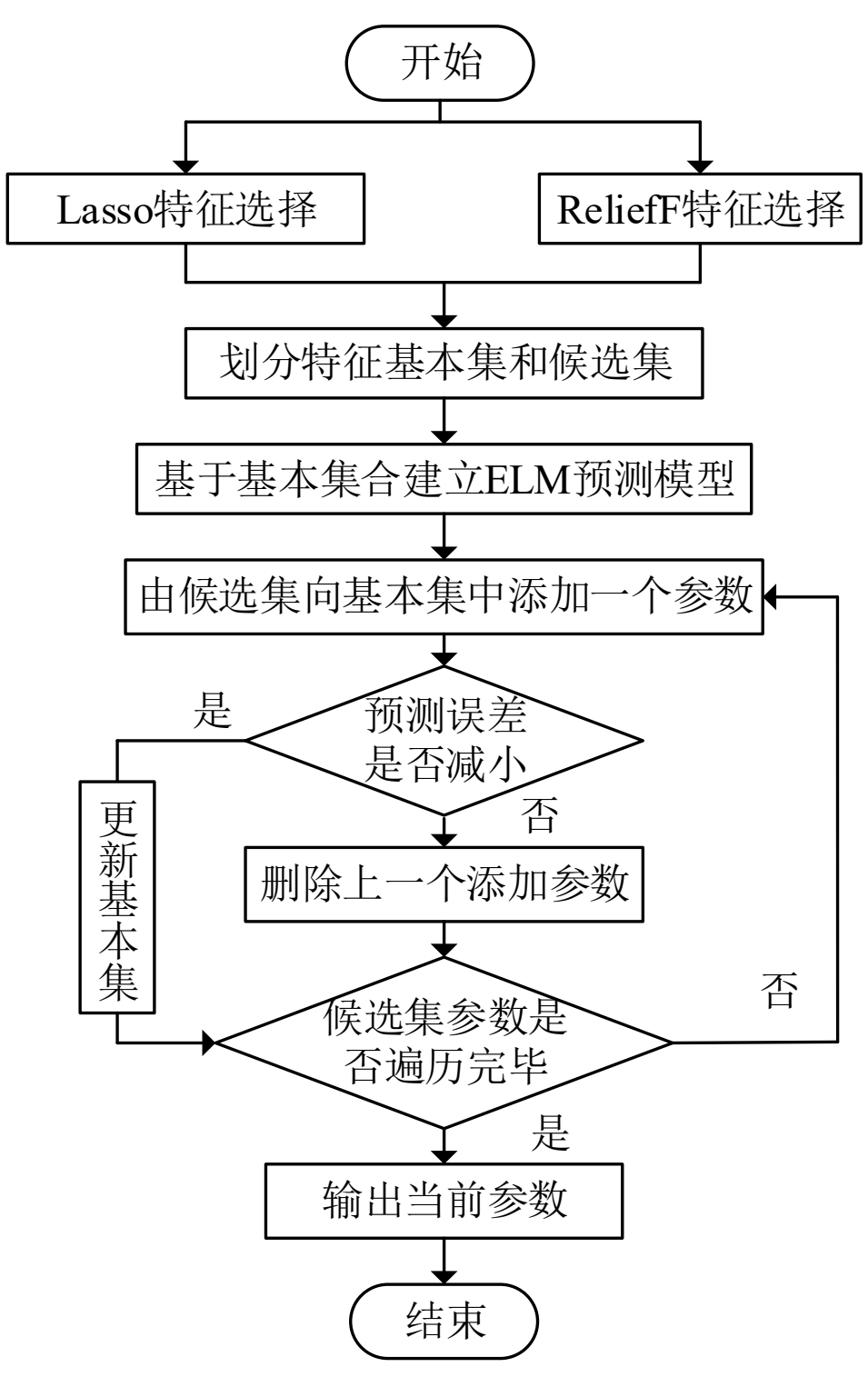

**图 2 自适应特征选择流程图**

**Fig. 2 Flow chart of adaptive feature selection**

Step1：结合锅炉燃烧反应过程的机理分析，确定预测模型的初始候选特征集合，然后使用 Lasso 和 ReliefF 对全部候选特征进行重要性排序。

Step2：将在两种算法排序中均排名前 50%的公共变量作为特征基本集合，其余变量作为候选集合；

Step3：向基本集中依次添加候选集中的变量，进行建模预测，记录测试结果；

Step4：判断更新基本集后的模型预测误差是否减小，如果减小，转步骤 5；否则，转步骤 6；

Step5：保留该变量到基本集合中，更新基本集合，转步骤 7；

Step6：删除上一个添加参数，转步骤 7；

Step7：判断候选集参数是否遍历完毕，如果完毕，转步骤 8；否则，转步骤 3；

Step8：输出最终更新完毕的特征基本集合。

使用上述方法针对 NOx 排放量进行特征选择，三种工况下的选择结果如图 3 所示：

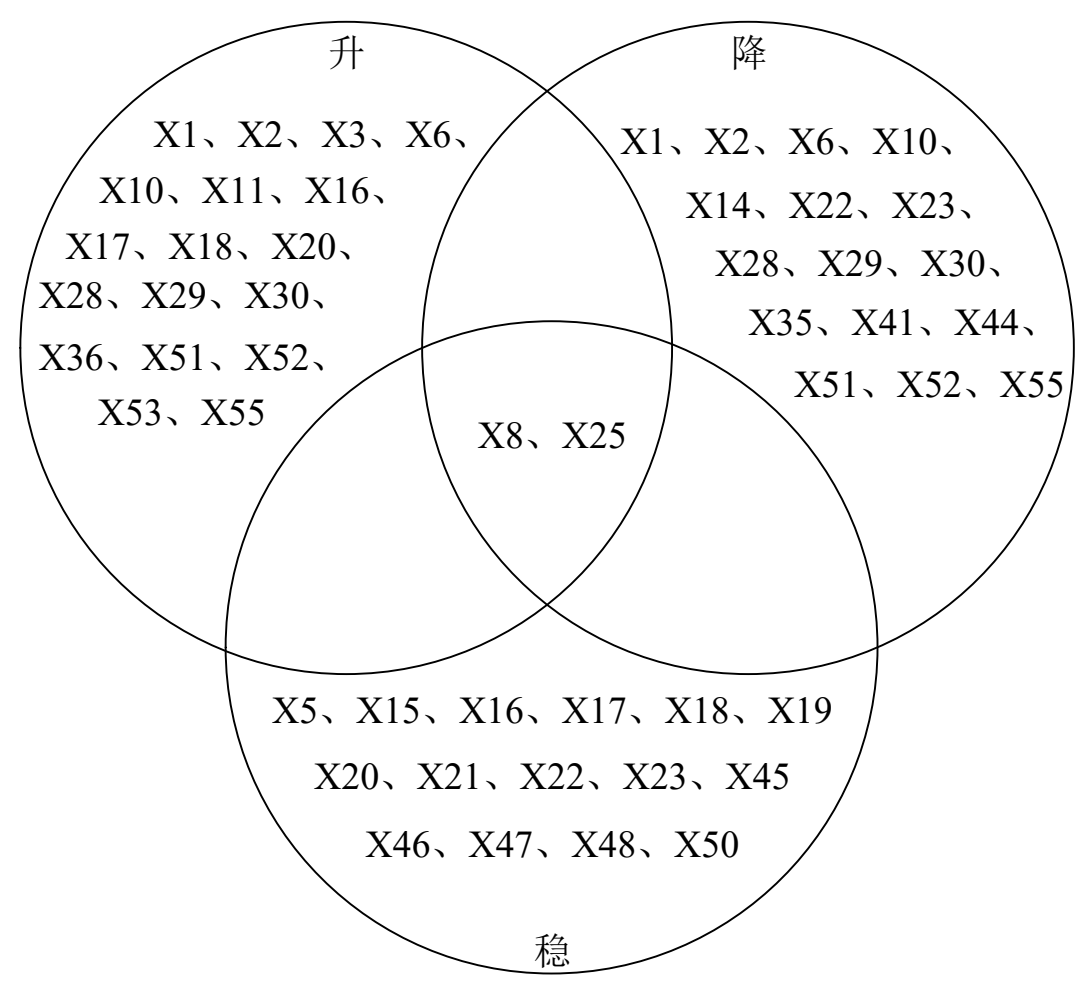

**图 3 不同工况建模使用的变量**

**Fig. 3 variables used in modeling under different conditions**

结合自适应特征选择结果及锅炉现场运行经验可得，不同运行工况下影响燃烧过程氮氧化物生成量的特征变量不同。现场锅炉实际运行中，影响锅炉运行的参数种类众多，其中相同类型输入变量存在较强相关性，例如每层四周的二次风量数据的变化趋势接近，因此在特征选择时对于同类的多个输入变量只选择其中部分变量。在机组稳态运行时，机组二次风配风方式及总煤量等主要影响氮氧化物生成量，因此特征选择算法将其选为输入特征变量；机组负荷上升或下降时，由于锅炉运行状态的改变，影响氮氧化物生成的变量的重要性发生改变，自适应特征选择算法的结果与稳态运行时的结果具有较大差异。因此，需要设计自适应特征选择算法挖掘锅炉不同运行工况下影响氮氧化物生成的不同特征集合。

## 2.4 基于 ELM 的预测模型

ELM[24]的网络结构为单隐层前馈神经网络(single-hidden-layer feedforward networks，SLFN)，采用一种新型的快速学习算法，具有学习快速、良好的泛化性和参数设置简单等优点，被广泛应用于各个领域。其网络结构如图 4 所示[25]。设训练样本为 $\{(x_t, y_t)\}_{t=1}^{N}$，其中样本总数为 $N$，第 $t$ 个样本输入向量为 $x_t = [x_{t1}, x_{t2}, \cdots x_{tn}]^{\mathrm{T}}$，$x_t$ 的维度为 $n$，输出向量为 $y_t = [y_{t1}, y_{t2}, \cdots y_{tm}]^{\mathrm{T}}$，$y_t$ 的维度为 $m$，则一个有 $L$ 个隐层节点的 ELM 可以表示为[26]：

$$\sum_{i=1}^{L} \beta_i f(\omega_i \cdot x_t + b_i) = y_t \tag{7}$$

其中，$f(x)$ 为激活函数，$\omega_{ij}$ 为输入权重，$\beta_i$ 为输出权重，$b_i$ 是第 $i$ 个隐层单元的偏置。$y$ 为网络输出向量。

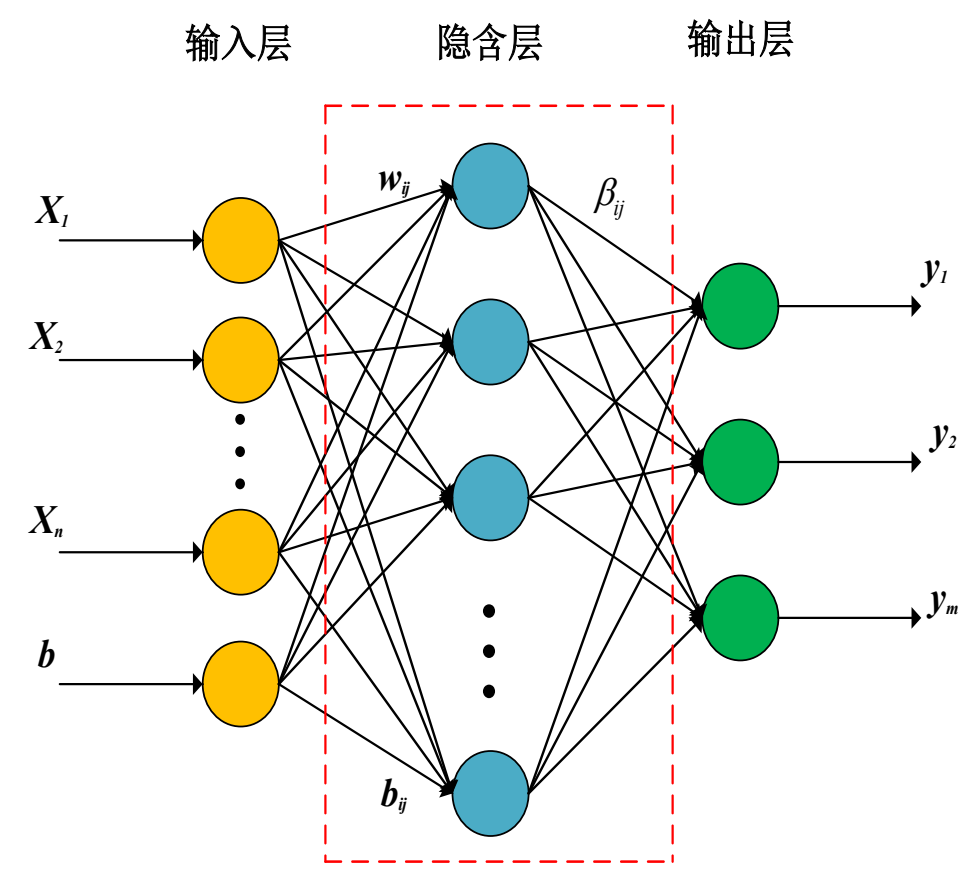

图 4　ELM 网络结构图

**Fig. 4　ELM network structure diagram**

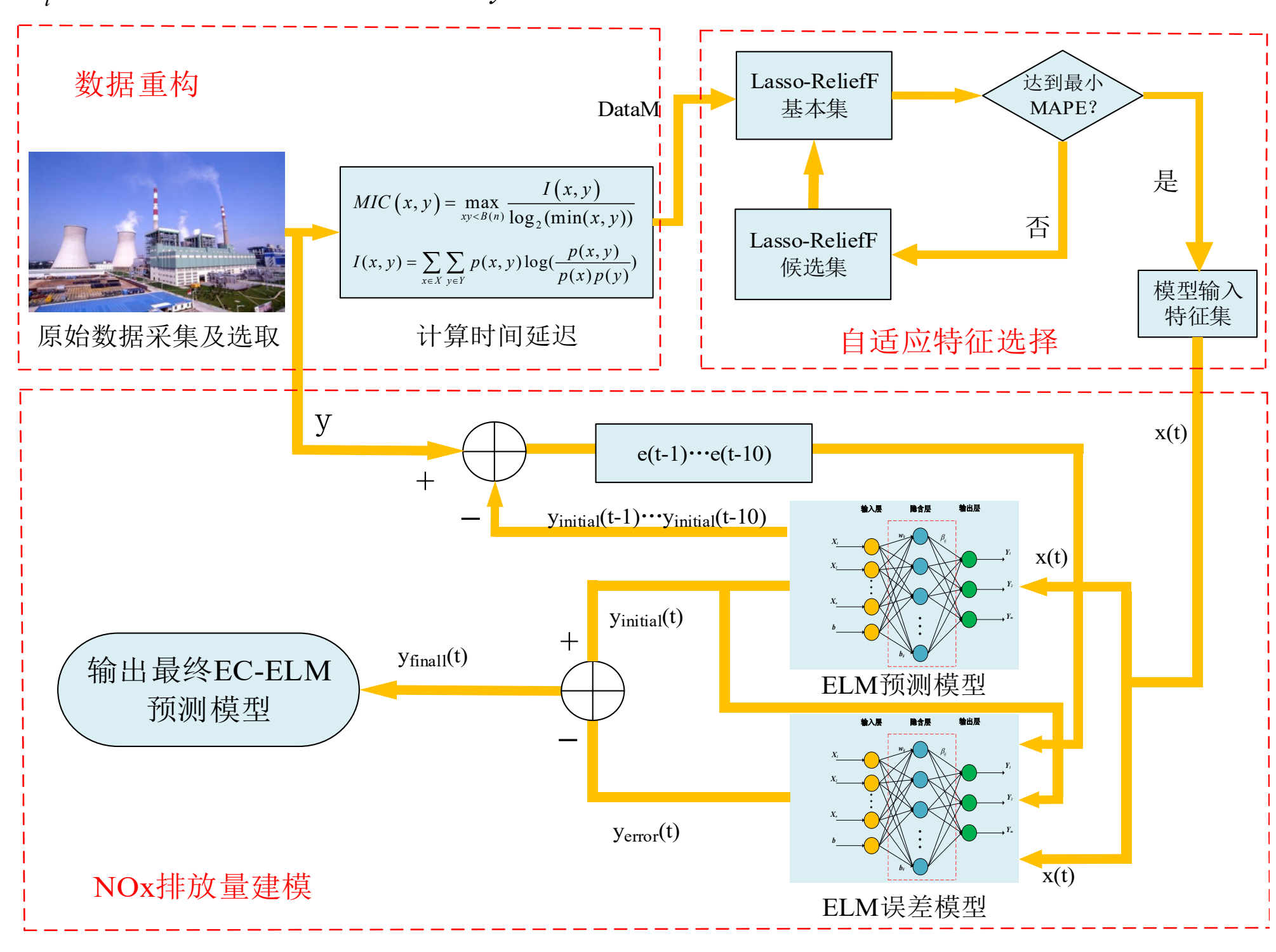

图 5　EC-ELM 模型结构图

**Fig. 5　Diagram of EC-ELM mode**

## 2.5 基于 ELM 的误差模型

由于模型在较长时间的预测过程中，机组运行工况会随时间发生改变，可能导致预测样本超出初始建模数据的分布范围，因此产生预测精度降低的情况。为使预测模型具备长期高精度预测能力，提出一种动态误差修正的预测方法。首先，公式(8)和(9)定义了预测模型的误差。通过 ELM 预测模型将训练集的预测值与测量值作差，得到的结果为误差 $e(t)$，将 $e(t)$ 作为误差模型的输出，并将前 10 时刻的历史误差 $e(t-1)\cdots e(t-10)$ 及原始模型的输入作为误差模型的输入变量，经训练得到误差模型，将误差模型的测试集输出与预测模型的测试集输出相加得到最终预测结果，即为误差修正 ELM（error-corrected ELM，EC-ELM）

模型，EC-ELM 模型结构如图 5 所示。

$$e(t)=y(t)-y_{initial}(t) \tag{8}$$

$$y_{finall}(t)=y_{initial}(t)+y_{error}(t) \tag{9}$$

式中：$e_t$ 为预测误差，$y(t)$ 为 NOx 测量值，$y_{initial}(t)$ 为 NOx 初始预测值，$y_{error}(t)$ 为误差预测值，$y_{finall}(t)$ 为 NOx 最终预测值。

## 3 实验结果及分析

使用平均绝对百分比误差(MAPE)、平均绝对误差(MAE)、归一化均方差(NMSE)、相关系数($R^2$)四个评价指标，来作为模型预测结果的评价标准，其公式计算分别如(10)-(13)所示[27]：

$$MAPE=\frac{1}{N}[\sum_{i=1}^{N}\frac{\left|Y_i-\hat{Y}_i\right|}{Y_i}]\times 100\% \tag{10}$$

$$MAE=\frac{1}{N}\sum_{i}^{N}\left|Y_i-\hat{Y}_i\right| \tag{11}$$

$$NMSE=\frac{1}{N}\sum_{i=1}^{N}\frac{(\hat{Y}_i-Y_i)^2}{Y_i\cdot\hat{Y}_i} \tag{12}$$

$$R^2=1-\frac{\sum_{i}(\hat{Y}_i-Y_i)^2}{\sum_{i}(\bar{Y}_i-Y_i)^2} \tag{13}$$

其中 $N$ 为测试集样本的个数，$Y_i$ 为实际测量值，$\hat{Y}_i$ 为模型预测值，$\bar{Y}_i$ 为样本均值。

为了验证该方法的有效性，实验应该采取更多的负载变化及运行条件，从而使模型适应在变工况条件下可变操作。变工况数据集包含三个工况，即锅炉的增负荷、降负荷和稳态负荷。稳态负荷的数据集共 4700 组，前 3500 组作为训练数据集，后 1200 组作为测试数据集。增负荷和降负荷的数据集分别有 4500 组和 3900 组数据，增负荷数据前 3500 组作为训练数据集，后 1000 组作为测试数据集。降负荷数据前 3000 组作为训练数据，后 900 组作为测试数据集。将稳态负荷、降负荷、增负荷这三组不同工况下的数据集分别用 D1、D2、D3 来表示。本实验采用 ELM、RBF、ESN、RBF 作为对比模型，已对所有对比模型进行超参数的网格搜索调试，以确保最佳的性能，保证各模型之间的公平性。

### 3.1 时间延迟对预测结果的影响

在燃烧过程中，由于锅炉燃烧系统具有较大的时延特性，延迟可能导致较低的预测精度。对于时间延迟，时序的调整对改善模型的预测性能至关重要。将各工况特征选取后所选择的变量分别进行有时延和无时延的处理，使用不同模型分别对各工况在有时延和无时延处理后的数据集进行验证。表 3 为不同工况下两种处理方法的评价指标，以 D2 为例，其中各预测模型的 MAPE 较未进行时间延迟处理的预测结果分别降低了 20.9%、6.9%、37.2%、12.8%，$R^2$ 分别提高了 8.2%、1.9%、61.7%、11%，MAE 和 NMSE 也都明显好于无时间延迟处理的预测结果。为了进一步看出特征选择对模型预测精度的影响，将各工况预测结果的误差通过箱型图进行展示，各工况误差结果的箱型图如图 6 所示。结果表明，经过时间延迟处理后的预测结果明显好于未进行时间延迟处理的预测结果。

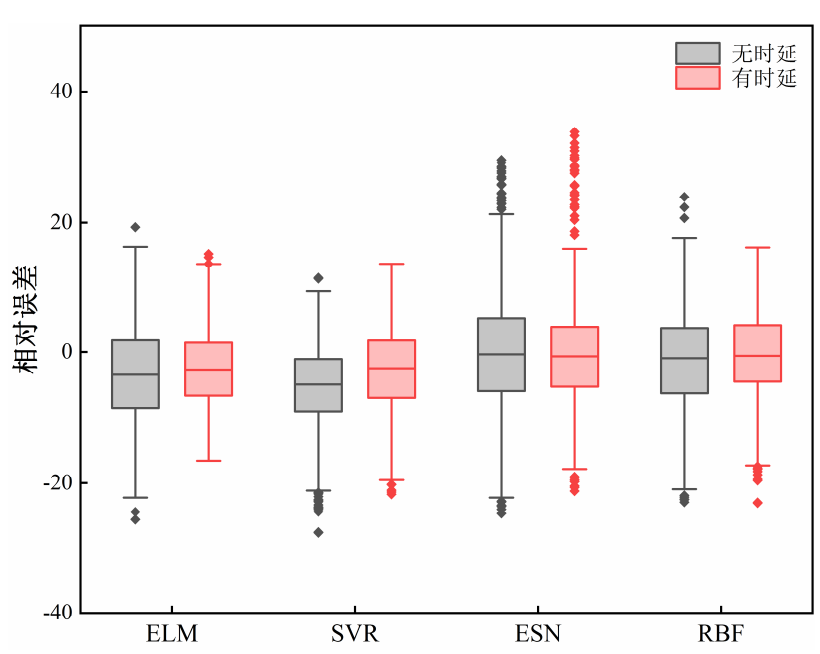

(a) D1 数据集

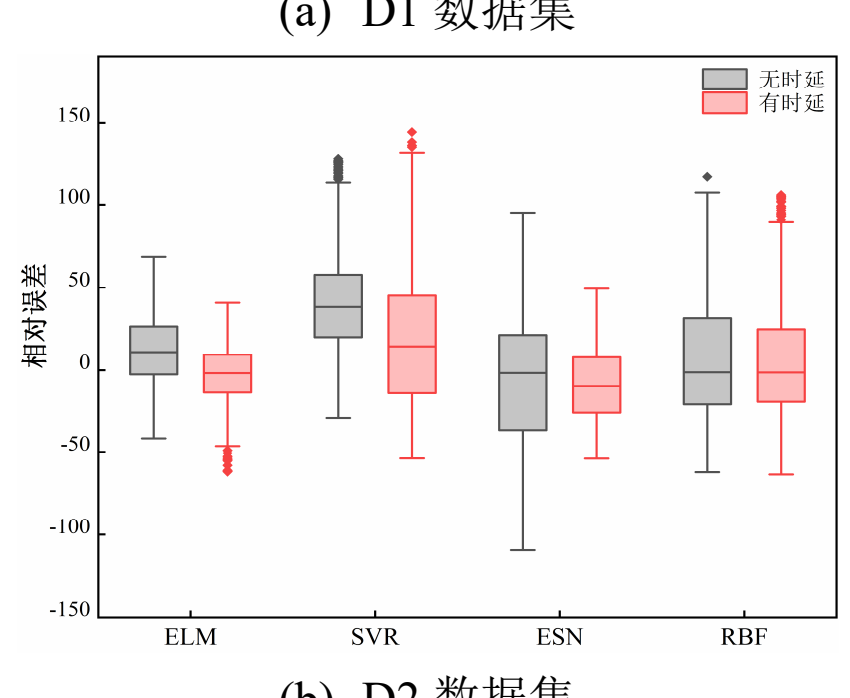

(b) D2 数据集

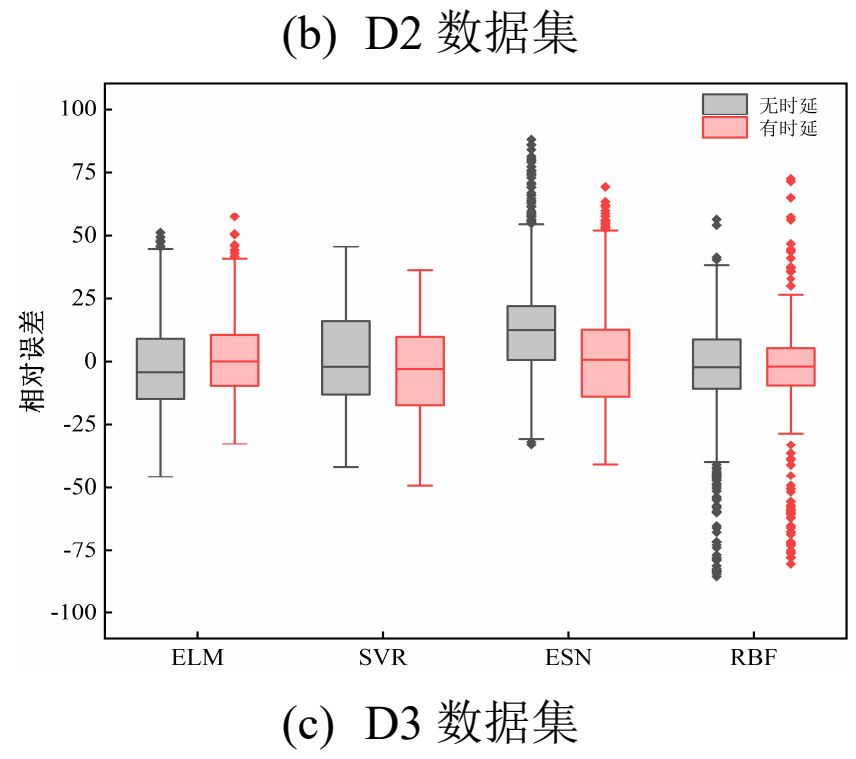

(c) D3 数据集

**图 6 时延前后误差箱型图**

**Fig. 6 Error box of prediction results before and after time delay**

表 3 不同模型有无时间延迟实验结果对比

**Tab. 3 Comparison of experimental results of different models with or without time delay**

| Different conditions | Evaluating indicator | ELM 是 | ELM 否 | SVR 是 | SVR 否 | ESN 是 | ESN 否 | RBF 是 | RBF 否 |
|---|---|---|---|---|---|---|---|---|---|
| D1 | MAPE | 2.60 | 3.16 | 2.73 | 3.07 | 2.68 | 3.44 | 2.47 | 2.86 |
| | $R^2$ | 0.68 | 0.55 | 0.63 | 0.56 | 0.59 | 0.40 | 0.70 | 0.61 |
| | MAE | 5.31 | 6.47 | 5.53 | 6.24 | 5.53 | 7.04 | 5.08 | 5.86 |
| | NMSE | 0.0010 | 0.0015 | 0.0012 | 0.0015 | 0.0014 | 0.0021 | 0.0010 | 0.0013 |
| D2 | MAPE | 3.78 | 4.88 | 8.45 | 9.97 | 4.94 | 7.86 | 6.70 | 7.69 |
| | $R^2$ | 0.87 | 0.81 | 0.32 | 0.20 | 0.82 | 0.51 | 0.57 | 0.51 |
| | MAE | 15.29 | 19.44 | 34.62 | 41.19 | 19.36 | 30.46 | 27.35 | 30.73 |
| | NMSE | 0.0021 | 0.0038 | 0.0130 | 0.0156 | 0.0033 | 0.0105 | 0.0075 | 0.0098 |
| D3 | MAPE | 5.06 | 6.99 | 7.25 | 7.74 | 7.41 | 8.91 | 4.82 | 6.54 |
| | $R^2$ | 0.76 | 0.55 | 0.62 | 0.48 | 0.51 | 0.21 | 0.66 | 0.44 |
| | MAE | 11.11 | 15.04 | 15.07 | 16.57 | 15.80 | 18.26 | 10.73 | 14.52 |
| | NMSE | 0.0041 | 0.0081 | 0.0076 | 0.0086 | 0.0099 | 0.0241 | 0.0045 | 0.0074 |

### 3.2 特征选择对预测结果的影响

在 NOx 预测中，主要是选择良好的输入变量集。输入变量必须从与 NOx 产生的有关过程中选择。本节分析 Lasso-ReliefF 特征选取对 ELM 模型预测结果的影响。在都进行时间延迟处理后，分别对未进行特征选取和Lasso-ReliefF 特征选取得到的特征集使用 ELM 模型对 NOx 进行预测。通过 Lasso-ReliefF 的实验结果均好于未进行特征选取的实验结果。以数据集 D1 为例，ELM 在所有预测模型中表现出良好的预测性能，虽然 RBF 的预测模型在 D1 中的预测结果略优于 ELM，但在其他工况中其预测结果表现出较大的波动，尤其在 D3 中，虽然 RBF 预测模型的 MAPE 低于 ELM 预测模型，但 $R^2$ 却低于 ELM，$R^2$ 表明了预测值与测量值的相关性以及拟合趋势，这说明 ELM 预测模型的预测结果与测量值之间相关性更大。ELM 预测模型在所有工况中均表现出稳定的预测性能。其中 ELM 预测模型的 MAPE 在 D1、D2、D3 中分别降低了 17.1%、56.9%和 35.7%，其他预测模型的预测结果在特征选择之后也好于特征选择之前，不同模型预测结果的评价指标如表 4 所示。

从表 4 中可以看出通过 Lasso-ReliefF 特征提取的方法可以获得更好的预测结果，这避免了单一的特征提取忽略一些重要的变量也避免了未进行特征提取而引入过多的冗余变量。如图 7(a)(b)(c)(d)可以清楚看到测量值和预测值的拟合情况。

表 4 不同模型有无特征选择实验结果对比

**Tab. 4 Comparison of experimental results of different models with or without feature selection**

| Different conditions | Evaluating indicator | ELM 是 | ELM 否 | SVR 是 | SVR 否 | ESN 是 | ESN 否 | RBF 是 | RBF 否 |
|---|---|---|---|---|---|---|---|---|---|
| D1 | MAPE | 2.60 | 3.14 | 2.73 | 3.07 | 2.68 | 4.11 | 2.47 | 3.25 |
| | $R^2$ | 0.68 | 0.53 | 0.63 | 0.58 | 0.59 | 0.18 | 0.70 | 0.48 |
| | MAE | 5.31 | 6.46 | 5.53 | 6.28 | 5.53 | 8.48 | 5.08 | 6.66 |
| | NMSE | 0.0010 | 0.0014 | 0.0012 | 0.0013 | 0.0014 | 0.0025 | 0.0010 | 0.0017 |
| D2 | MAPE | 3.78 | 8.97 | 8.45 | 9.15 | 4.94 | 10.13 | 6.70 | 5.83 |
| | $R^2$ | 0.87 | 0.37 | 0.32 | 0.25 | 0.82 | 0.16 | 0.57 | 0.73 |
| | MAE | 15.29 | 35.47 | 34.62 | 38.28 | 19.36 | 40.64 | 27.35 | 22.76 |
| | NMSE | 0.0021 | 0.0106 | 0.0130 | 0.0135 | 0.0033 | 0.0156 | 0.0075 | 0.0052 |
| D3 | MAPE | 5.06 | 7.87 | 7.25 | 10.51 | 7.41 | 10.55 | 4.82 | 5.74 |
| | $R^2$ | 0.76 | 0.48 | 0.62 | 0.16 | 0.51 | 0.19 | 0.66 | 0.65 |
| | MAE | 11.11 | 16.82 | 15.07 | 21.39 | 15.80 | 22.46 | 10.73 | 12.36 |
| | NMSE | 0.0041 | 0.0092 | 0.0076 | 0.0252 | 0.0099 | 0.0184 | 0.0045 | 0.0053 |

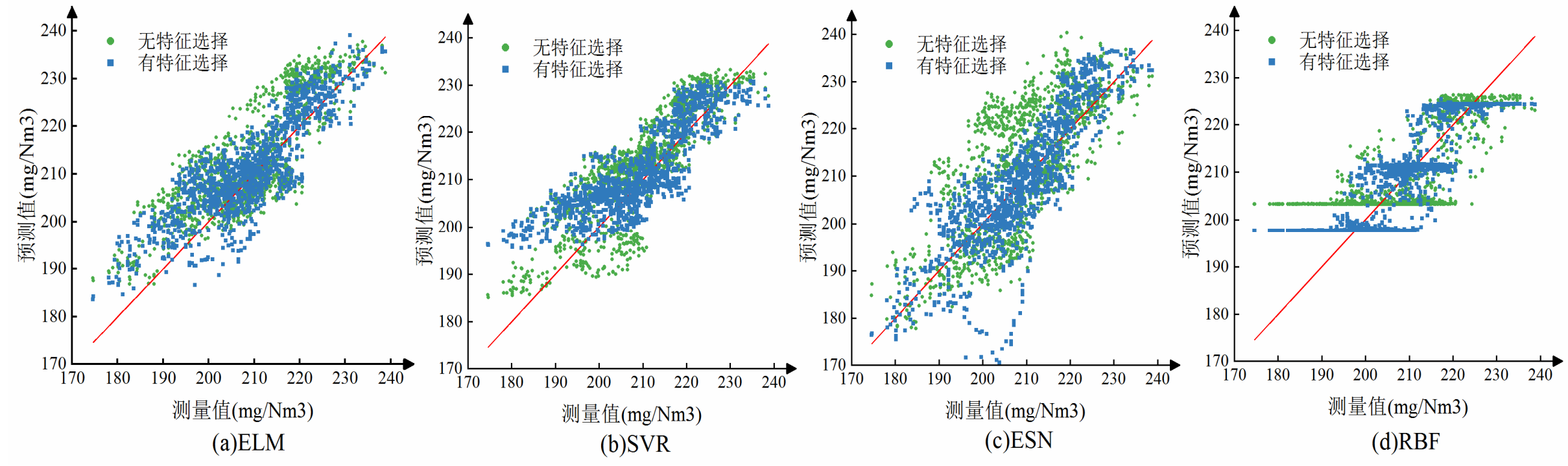

图 7　不同模型预测结果对比图（有/无特征选择）

**Fig. 7　Comparison of prediction results of different models with/without feature selection**

## 3.3 误差修正对预测结果的影响

为了验证本文所提出的误差修正策略的有效性，分别采用 ELM、ESN、RBF、SVR 这四种预测模型对有误差修正和没有误差修正的策略进行验证，并比较其结果。通过对各预测模型产生的误差进行误差分析，经过误差修正后的预测值，精度有了明显的提高，表明了误差修正策略的有效性。以数据集 D2 为例，如表 5 所示，其中 ELM 模型在所有模型中表现出最优的预测精度，在误差修正后，ELM 预测模型的 MAPE 降低了 58.7%，$R^2$ 提高了 11.8%，MAE 降低了 59.0%，NMSE 降低了 81%。为了更明显更直观的看出该策略的有效性，图 8 比较了不同策略的平均绝对百分比误差，从图 8 中可以看出 ELM 模型的误差修正后的结果优于没有误差修正的结果。其他预测模型在误差修正后所有评价指标也都有了明显的提高。这表明了误差修正这一策略的有效性，该策略不止对 ELM 预测模型有效，在其他预测模型中也一样适用。

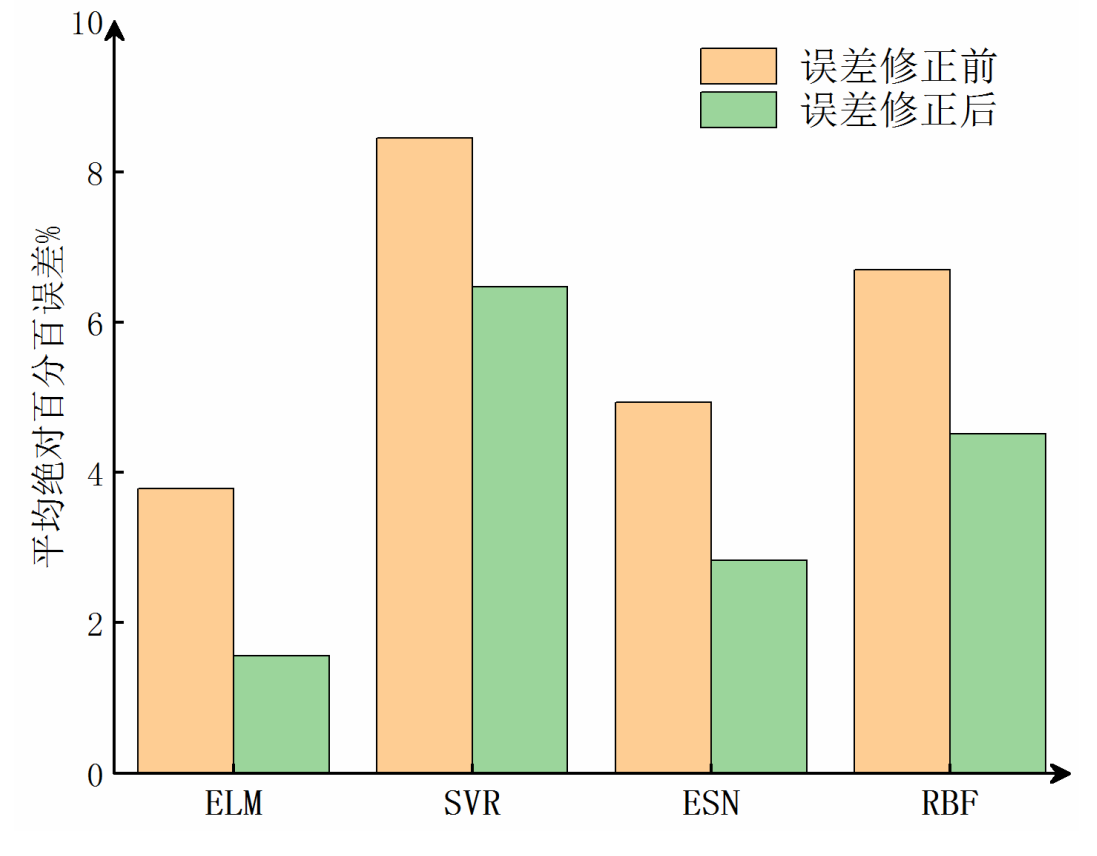

图 8　误差修正前后结果对比

**Fig. 8　Comparison of results before and after error correction**

表 5 误差修正前后模型误差比较

**Tab. 5　Comparison of model errors before and after error correction**

| 模型 | 方法 | MAPE/% | R2 | MAE | NMSE |
|---|---|---|---|---|---|
| ELM | 修正前 | 3.78 | 0.87 | 15.29 | 0.0021 |
| | 修正后 | 1.56 | 0.98 | 6.27 | 0.0004 |
| SVR | 修正前 | 8.45 | 0.32 | 34.62 | 0.0130 |
| | 修正后 | 6.47 | 0.69 | 25.75 | 0.0064 |
| ESN | 修正前 | 4.94 | 0.82 | 19.36 | 0.0033 |
| | 修正后 | 2.83 | 0.93 | 11.29 | 0.0014 |
| RBF | 修正前 | 6.70 | 0.57 | 27.35 | 0.0075 |
| | 修正后 | 4.51 | 0.78 | 18.10 | 0.0038 |

## 3.4 不同模型的对比

为了验证该模型的有效性，将 EC-ELM 模型与 ELM、SVR、ESN、RBF 五种模型进行预测结果对比。图 9 是不同模型的真实值与预测值的对比曲线图，实验结果表明 EC-ELM 模型能够更好的跟随实际值的变化，为了可以更好的看出各个模型的预测精度，从图中截取 200-300 这部分曲线用于更好的观察拟合程度，可以看出 EC-ELM 模型的真实值与预测值较其他几种模型的拟合曲线最优，这说明 EC-ELM 模型可以更有效的预测 NOx 排放浓度。

为了进一步展示出 EC-ELM 模型的预测性能，图 10 比较了不同模型的绝对误差频数分布情况。从图 10 中可以看出，在不同工况下，EC-ELM 模型的绝对误差大部分分布在[0,5]之间，其频率随着绝对误差的升高而逐渐降低，在高误差区域基本没有分布，其他几种模型也呈现和 EC-ELM 模型一样的趋势，但 EC-ELM 模型下降的最快，这表明了 EC-ELM 模型的预测精度远高于其他几种模型。从图 10(a)(b)(c)中可以得出，EC-ELM 模型在不同工况下都可以表现出良好预测性能，

表 6　不同工况的预测结果对比

**Tab. 6 Comparison of prediction results under different working conditions**

| Data sets | Evaluating indicator | EC-ELM | ELM | SVR | ESN | RBF |
|---|---|---|---|---|---|---|
| D1 | MAPE | 1.03 | 2.60 | 2.73 | 2.68 | 2.47 |
| | $R^2$ | 0.95 | 0.68 | 0.63 | 0.59 | 0.70 |
| | MAE | 2.14 | 5.31 | 5.53 | 5.53 | 5.08 |
| | NMSE | 0.0002 | 0.0010 | 0.0012 | 0.0014 | 0.0010 |
| D2 | MAPE | 1.56 | 3.78 | 8.45 | 4.94 | 6.70 |
| | $R^2$ | 0.98 | 0.87 | 0.32 | 0.82 | 0.57 |
| | MAE | 6.27 | 15.29 | 34.62 | 19.36 | 27.35 |
| | NMSE | 0.0004 | 0.0021 | 0.0130 | 0.0033 | 0.0075 |
| D3 | MAPE | 1.87 | 5.07 | 7.25 | 7.41 | 4.82 |
| | $R^2$ | 0.97 | 0.76 | 0.62 | 0.51 | 0.66 |
| | MAE | 3.98 | 11.11 | 15.07 | 15.80 | 10.73 |
| | NMSE | 0.0005 | 0.0041 | 0.0076 | 0.0099 | 0.0045 |

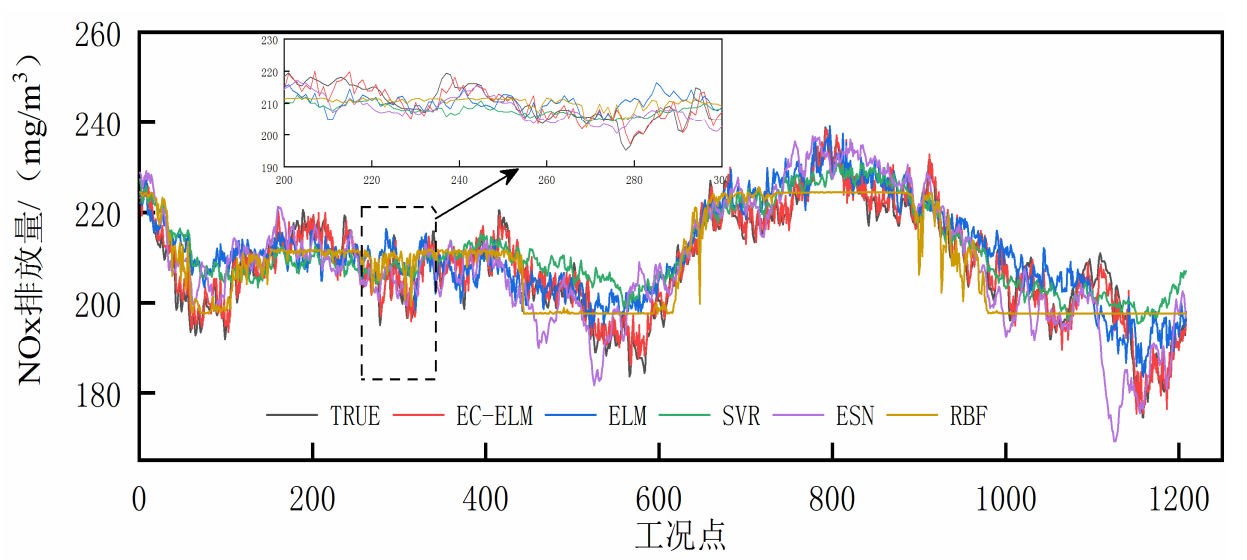

(a)　D1 data sets

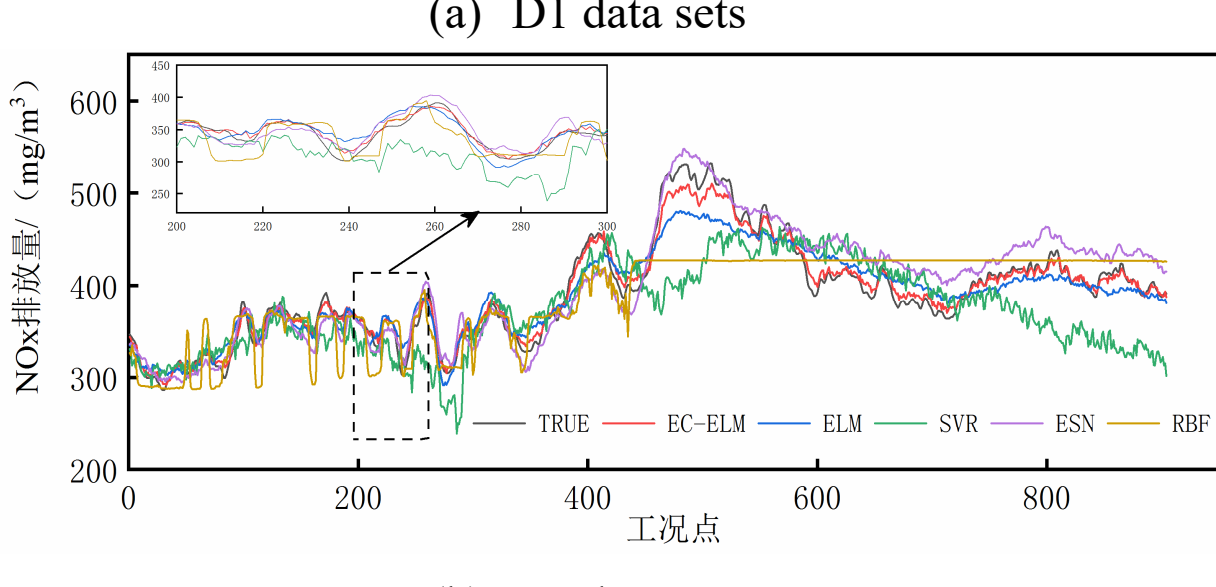

(b)　D2 data sets

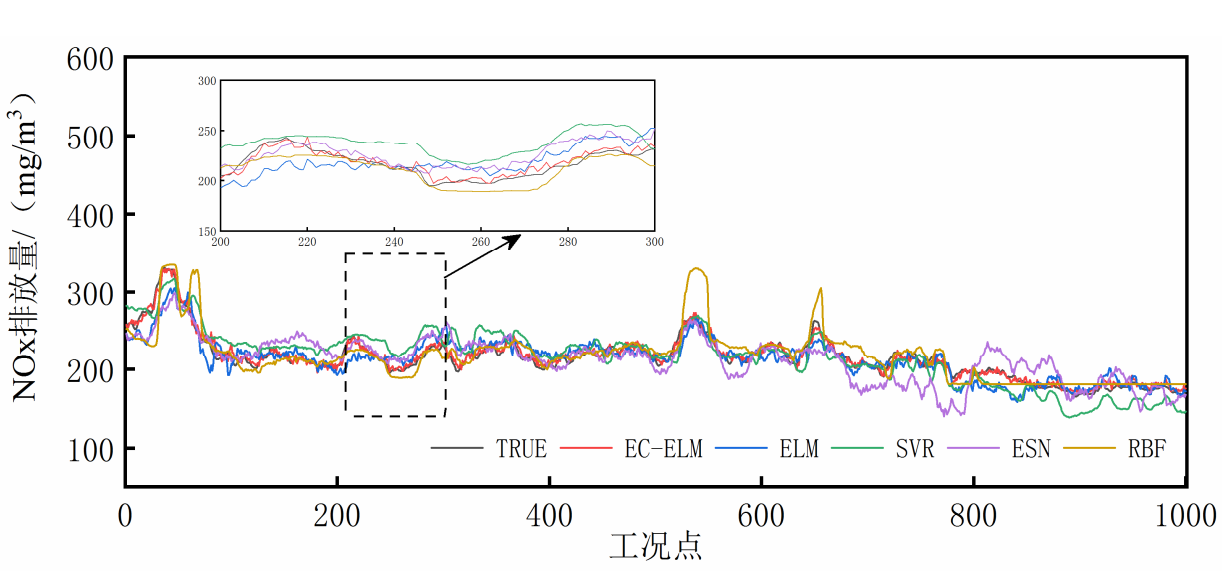

(c)　D3 data sets

图 9　不同模型的预测结果

**Fig. 9　Prediction results of different models**

这也进一步体现了 EC-ELM 模型的稳定性。

通过几种评价指标将几种模型进行对比，表 6 是不同工况下五种模型的预测结果。在所有负

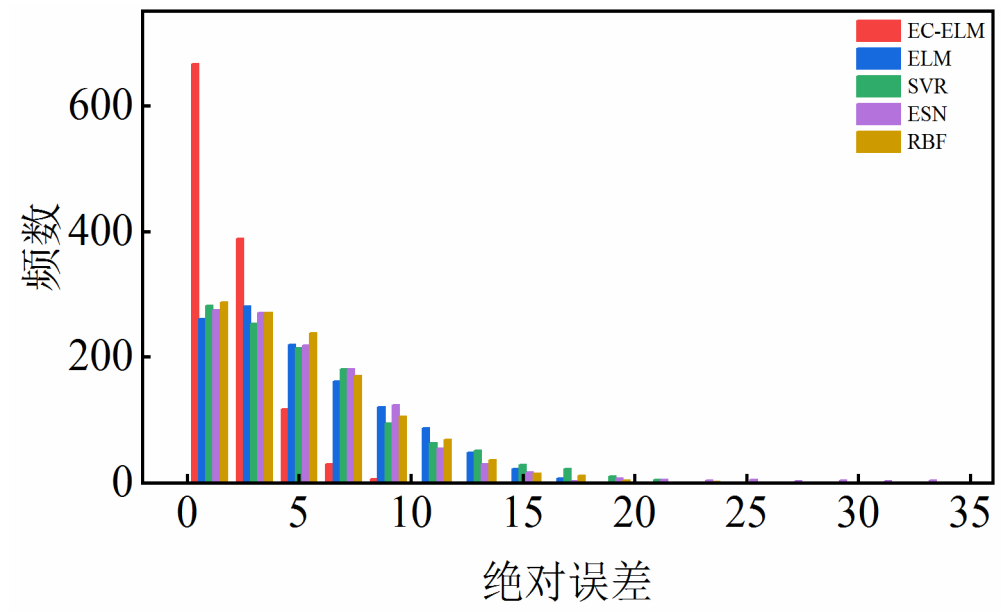

(a)　D1 data sets

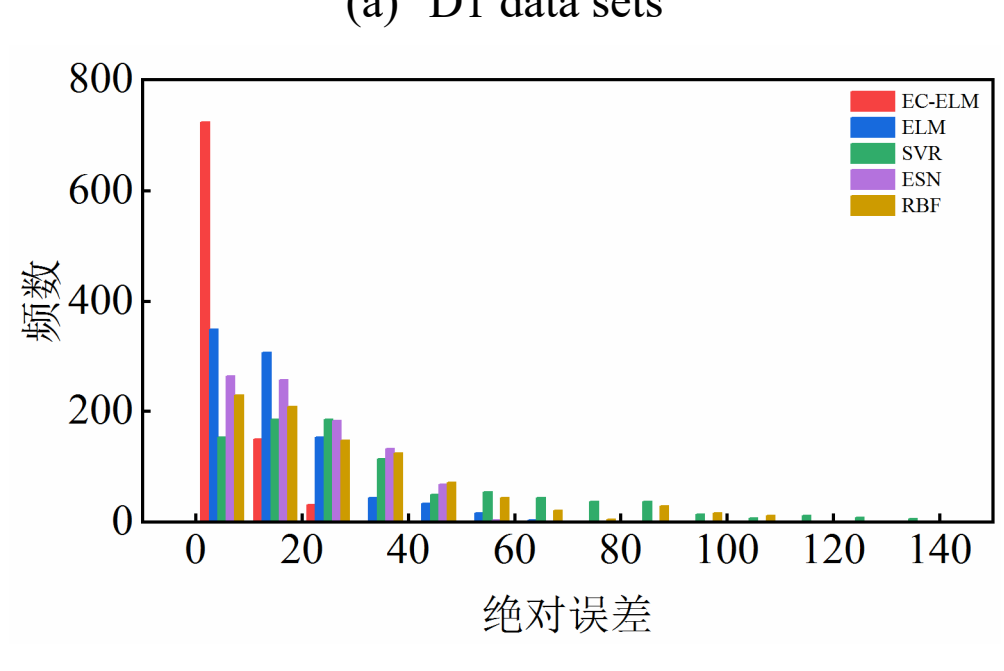

(b)　D2 data sets

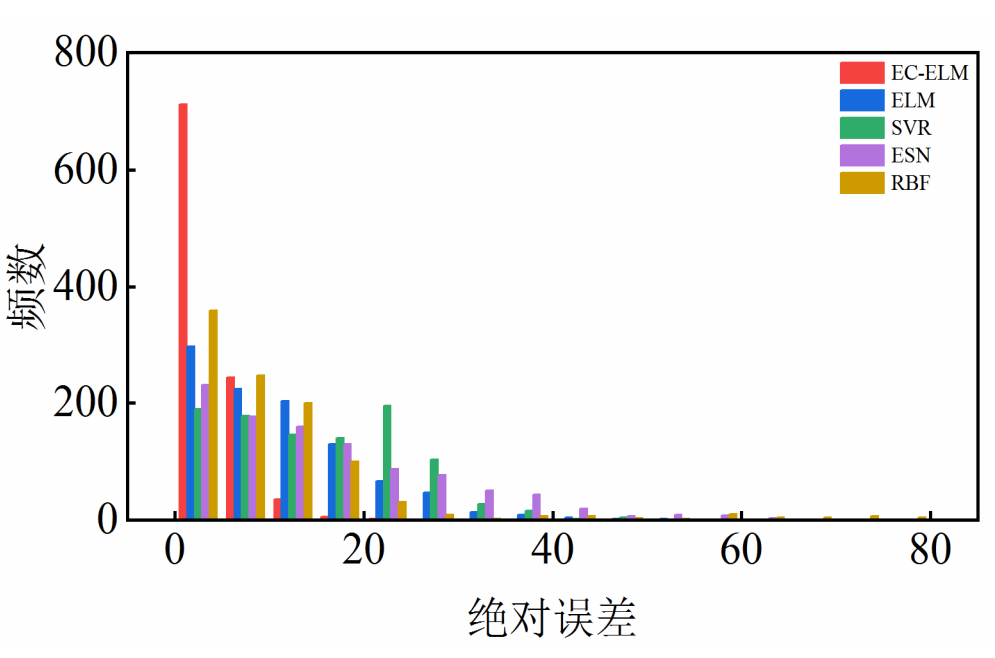

(c)　D3 data sets

图 10　不同模型的绝对误差

**Fig. 10　Absolute error of different models**

荷条件下，EC-ELM 模型的 MAPE 明显好于其他四种模型，并且在五种模型对比中，EC-ELM 模型的其他各项评价指标均优于其他三种模型。在数据集 D1 中，所提出算法的 MAPE 与其他几种算法相比分别降低了 60.4%、62.5%、61.5%、58.3%。在数据集 D2 中，MAPE 分别降低了 58.7%、81.5%、68.4%、76.7%。在数据集 D3 中 MAPE 分别降低了 63.1%、74.2%、74.8%、61.2%。实验结果证明，EC-ELM 模型与其他几个模型相比较具有更好的预测性能。

### 3.5 讨论

通过实验结果分析得到以下讨论：

1)锅炉是一个具有大时滞的燃烧系统，由于传感器测量的延迟以及各项燃烧反应的时间，各输入参数与输出之间存在一定的时间延迟，通过计算各参数的时间延迟，进行建模数据的重构，提升建模数据之间的非线性相关性。实验结果表明该计算延迟时间方法的有效性；

2)Lasso-ReliefF 自适应特征选择方法结合了 Lasso 和 ReliefF 两种算法，充分考虑各参数对于建模精度的影响，避免漏选重要性高的输入参数，同时减少冗余变量的输入，因此，提升了模型的预测精度；

3)通过初始预测模型的误差建立动态误差校正模型，对初始预测模型的预测值进行校正，从而进一步降低 NOx 排放量的预测误差。实验结果表明，经过误差修正后，模型预测的精度得到了提高；

4)通过在不同工况下 EC-ELM 模型与其他几种模型的比较讨论，证明了 EC-ELM 模型在不同工况 NOx 排放预测问题的精确性、稳定性和适用性。

## 4 结论

针对燃煤机组 NOx 排放机理复杂且 CEMS 系统存在工作停滞的问题，提出了一种基于数据驱动的燃煤机组 NOx 排放量动态修正预测模型，采用历史运行数据进行试验，得到以下结论：

1)基于最大信息系数 MIC 计算相关参数与 NOx 排放量之间的延迟时间，消除燃烧过程延迟特性对建模精度的影响，提升了模型的预测精度；

2)设计自适应特征选择策略，确定对 NOx 排放量影响较大的参数作为模型输入参数，使预测模型在三种数据集上的预测精度均得到提升；

3)通过动态误差修正策略，对当前时刻 NOx 排放量的初始预测值进行误差校正，进一步提升预测精度。

所提方法可用于燃煤机组 NOx 排放量的精准动态预测，可作为机组燃烧过程优化调整及脱硝系统精准控制的基础。

## 参考文献

# Data Driven based Dynamic Correction Prediction Model for NOx Emission of Coal Fired Boiler

Tang Zhenhao[1*], Zhu Deyu[1], Li Yang[2]

(1. School of Automation Engineering, Northeast Electric Power University, 2. School of Electrical Engineering, Northeast Electric Power University)

CEMS measures the NOx emission of the coal-fired unit. However, due to the harsh working environment, CEMS sometimes has a work stoppage situation. Therefore, establishing the NOx prediction model has practical engineering significance. And the NOx emission prediction model is the cornerstone of combustion optimization and ammonia injection volume control. In this paper, a data-driven-based dynamic correction prediction model for NOx emission of coal-fired boiler is proposed.

The main structure of the proposed method is shown in Fig. 1. The modeling steps are as follows:

Step1: Obtain primary auxiliary variables through mechanism analysis.

Step2: Calculate the delay time between each auxiliary variable and the NOx emission and reconstruct the modeling data.

Step3: Obtain the final input variables of the prediction model through the Lasso-ReliefF adaptive feature selection algorithm.

Step4: Use ELM algorithm to establish NOx prediction model and error model.

Step5: Establish the final EC-ELM prediction model.

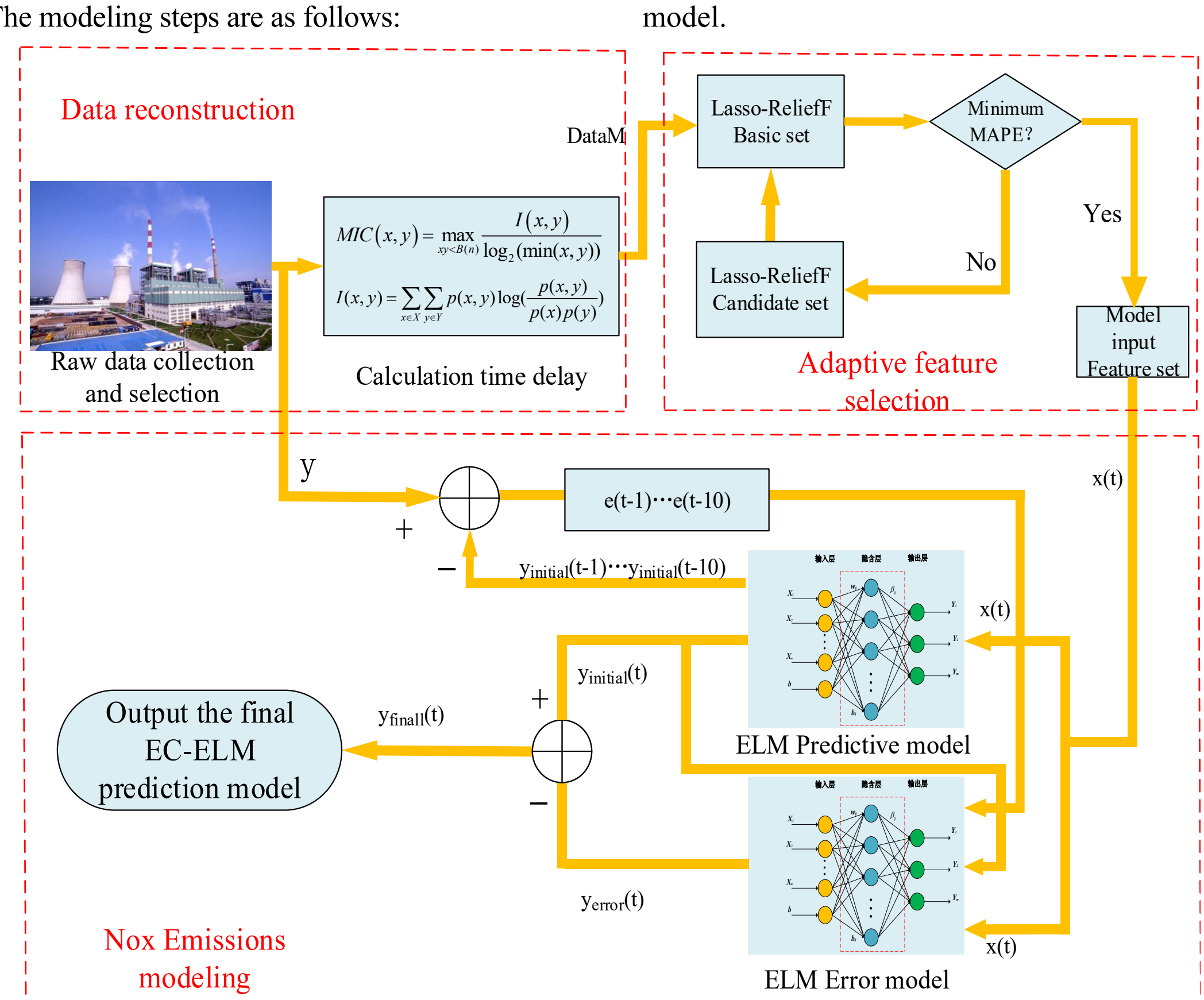

**Fig. 1 structure diagram of EC-ELM model**

The experimental results show that the method proposed in this paper can effectively predict NOx emissions, and the predicted value can follow the changes of the measured value well under different working conditions. The strategy is also applicable in different models.

In this abstract, Figure 1 is on page 6 of the original.